\documentclass[journal]{IEEEtran}
\newcommand{\ie}{\emph{i.e.}, }

\usepackage[utf8]{inputenc}
\DeclareUnicodeCharacter{200E}{}

\ifCLASSINFOpdf
   \usepackage[pdftex]{graphicx}
   \graphicspath{{../pdf/}{../jpeg/}}
   \DeclareGraphicsExtensions{.pdf,.jpeg,.png}
\else
\fi
\usepackage{amsmath}
\usepackage{multirow}
\usepackage{xcolor}
\usepackage{amsfonts}
\usepackage{cite}

\usepackage{algorithmic}
\usepackage[ruled,vlined]{algorithm2e}

\usepackage{array}

\ifCLASSOPTIONcompsoc
 \usepackage[caption=false,font=normalsize,labelfont=sf,textfont=sf]{subfig}
\else
 \usepackage[caption=false,font=footnotesize]{subfig}
\fi
\usepackage{fixltx2e}
\usepackage{url}

\begin{document}
%
\title{A Cascaded Approach for ultraly High Performance Lesion Detection and False Positive Removal in Liver CT Scans}


%
%

 
\author{Fakai Wang, Chi-Tung Cheng, Chien-Wei Peng, Ke Yan, \\Min Wu, \IEEEmembership{Fellow, IEEE}, Le Lu, \IEEEmembership{Fellow, IEEE}, Chien-Hung Liao, and Ling Zhang

\thanks{Fakai Wang and Min Wu are with the ECE Department, University of Maryland, College Park, MD, USA. (email:jackwangumd@gmail.com). Chi-Tung Cheng (email:atong89130@gmail.com) and Chien-Hung Liao are with the Department of Trauma and Emergency Surgery, Chang Gung Memorial Hospital at Linkou, Chang Gung University, Linkou, Taiwan, ROC. Chien-Wei Peng is with Department of Gastroenterology and Hepatology, Chang Gung Memorial Hospital, Linkou Medical Center, Linkou, Taiwan, ROC. Ke Yan, Le Lu, and Ling Zhang are with DAMO Academy, Alibaba Group.}
}

\maketitle

\begin{abstract}
Liver cancer has high morbidity and mortality rates in the world. Multi-phase CT is a main medical imaging modality for detecting/identifying and diagnosing liver tumors. Automatically detecting and classifying liver lesions in CT images have the potential to improve the clinical workflow. This task remains challenging due to liver lesions' large variations in size, appearance, image contrast, and the complexities of tumor types or subtypes. 
In this work, we customize a multi-object labeling tool for multi-phase CT images, which is used to curate a large-scale dataset containing 1,631 patients 
with four-phase CT images, multi-organ masks, and multi-lesion (six major types of liver lesions confirmed by pathology) masks. We develop a two-stage liver lesion detection pipeline, where the high-sensitivity detecting algorithms in the first stage discover as many lesion proposals as possible, and the lesion-reclassification algorithms in the second stage remove as many false alarms as possible. 
The multi-sensitivity lesion detection algorithm maximizes the information utilization of the individual probability maps of segmentation, and the lesion-shuffle augmentation effectively explores the texture contrast between lesions and the liver. 
Independently tested on 331 patient cases, the proposed model achieves high sensitivity and specificity for malignancy classification in the multi-phase contrast-enhanced CT (99.2\%, 97.1\%, diagnosis setting) and in the noncontrast CT (97.3\%, 95.7\%, screening setting). 


\end{abstract}


%
\IEEEpeerreviewmaketitle

\section{Introduction}
\IEEEPARstart{L}{iver} cancer ranks the 6th by incidence, but the 3rd by mortality in 2020~\cite{globalBurden2020}. Similar to other cancers, it is important to detect liver cancer in the early stage and take prompt action. Ultrasound and noncontrast computed tomography (CT) are usually used in the screening setting, while multi-phase contrast-enhanced CTs are used in the diagnosis setting. According to liver imaging reporting and data system (LI-RADS)~\cite{LIRADS} guidelines, contrast-enhanced CTs or MRIs are sufficient for identifying liver lesions in broad situations. Early detection and accurate diagnosis of liver lesions involve many challenges, such as low awareness, lack of examinations, missed detection of small lesions, and misclassification of lesion types. 

In the opportunistic screening setting (only the noncontrast [NC] CT scanning), small lesions usually have poor detection performance by radiologists, due to many factors, such as the complex contexts of the liver, ambiguous appearances of small lesions, and low image quality. Reading NC CT to detect potential liver lesions in the high-volume physical examination scenario where most patients have healthy livers is a tedious task, and the sensitivity of detection can be low. On the other hand, in the tumor diagnosis setting, reading multi-phase contrast-enhanced CT images (NC, arterial phase [AP], venous phase [VP], delayed phase [DP]) requires high expertise and costs great effort to compare local image patterns to localize lesions within the large liver organ, and then determine the lesion types. Therefore, developing algorithms that can automatically detect and classify liver lesions in both the screening and diagnosis setting is of great interest.

Developing highly accurate liver lesion detection and classification method for CT images face many challenges. The first challenge is inherent complex morphology. The liver is a large organ in the abdomen, and there are many morphology variations in both the liver and lesions, such as location, size, shape, intensity, and texture. The complex vessels and ducts in the liver and surrounding organs add to the difficulty of detecting and localizing lesions. Liver texture changes due to fat accumulation and fibrosis also create much ambiguity in lesion detection. There are many malignant and benign lesion types in the liver, and some types are hard to distinguish in the CT due to appearance similarities. 

The second challenge is the lack of ideal datasets; specifically, no large-scale, high-quality (e.g., covering various liver conditions, with labels of major tumor types, etc.) liver CT datasets are publicly available for deep learning models to learn the comprehensive patterns. According to the LI-RADS guideline, multi-phase contrast-enhanced liver CT images are needed for radiologists to determine the tumor types, and in some cases, pathology reports are required. The ideal labels are comprehensive voxel-level annotations of related organs, vessels, and liver lesions and, importantly, clinically-confirmed lesion types for all lesions, which are hard to obtain. Most existing studies rely on their established in-house datasets with weak labels, e.g., only box-level annotations of lesions and/or without labels of lesion types. In addition, the curation of a high-quality liver lesion dataset involves diagnosis-related privacy information and clinical practice in the medical center, making it harder to open source. 

The third challenge is how to design effective and clinically-relevant evaluation metrics for comprehensive performance judgment of the deep learning models. Currently, many works employ general image segmentation metrics, which are volume- and surface-based, such as Dice, average symmetrical surface distance (ASD), etc. Some works employ object detection metrics, such as recall, precision, and area under the curve (AUC). However, these general metrics do not adequately evaluate the model performance, not covering the detecting ability of small or suspected malignant lesions, which is clinically important in real-world applications. Therefore, we need to adopt fine-level evaluation metrics for individual lesions with varied characteristics. 

\begin{figure}[t]
	\centering
    \hfill
	\includegraphics[width=0.98\linewidth]{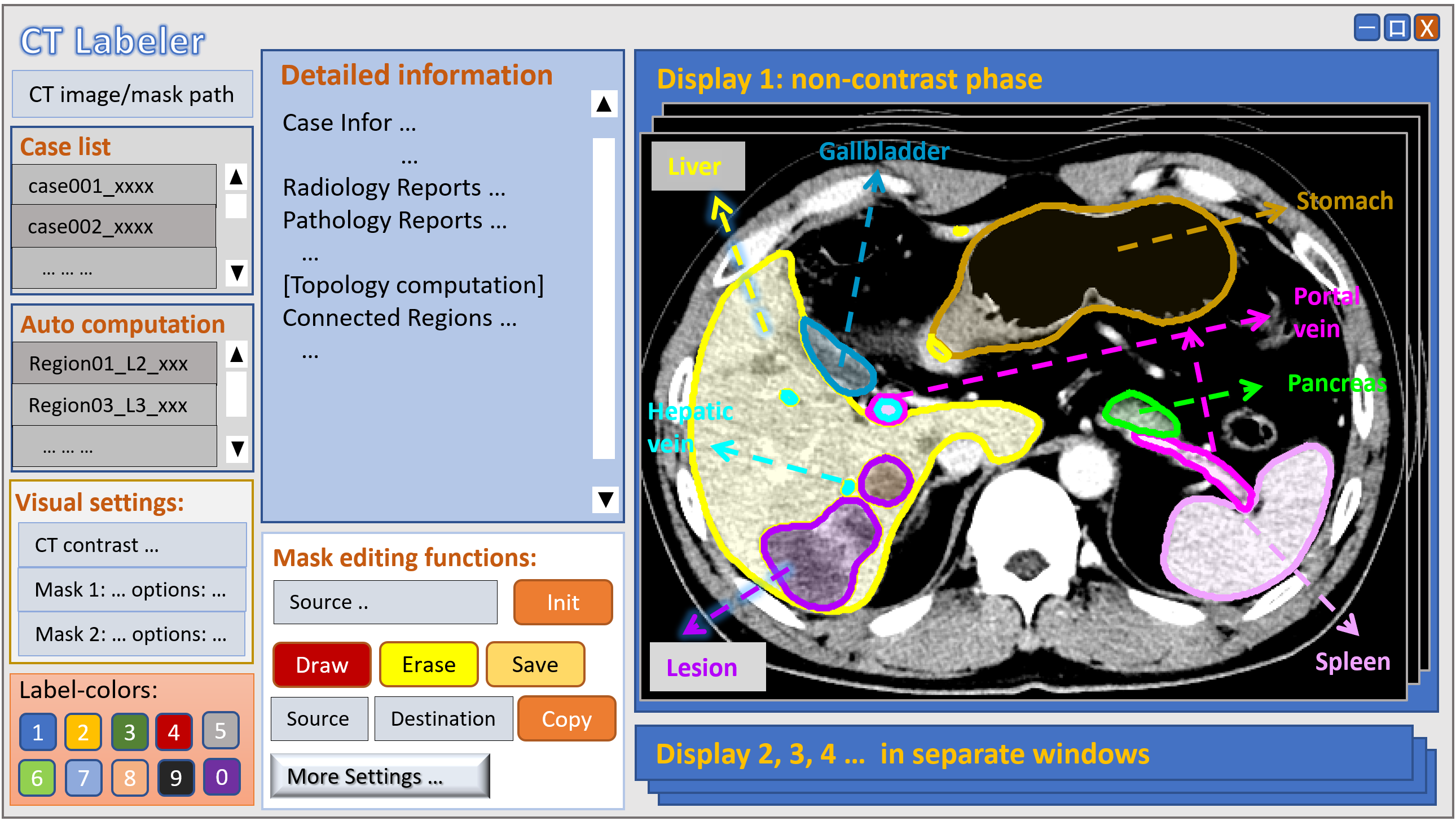}
	\hfill
	\caption{We propose solving the liver lesion detection and diagnosis task by a multi-class semantic segmentation. Besides segmenting lesions, we additionally include seven related organ\&vessel segmenting tasks for better lesion performance. We customize a multi-object labeling tool (CT Labeler) and adopt a semi-supervised learning approach to curate a large-scale multi-phase abdominal liver CT dataset with high-quality masks.}
	\label{fig:ctlabelerorgans}
\end{figure}

To address these challenges, we develop a customized multi-object labeling tool for multi-phase CT images (Fig. \ref{fig:ctlabelerorgans}), based on which we curate a large-scale (n=1,631) CT image dataset with organs\&vessels, and liver lesion masks. We further propose a multi-sensitivity segmentation algorithm for better lesion probabilistic inference, and introduce a lesion-shuffle training scheme for better small-lesion detection. Our contributions lie in four folds. First, to generate initial masks of organs\&vessels and liver lesions of our private dataset, we train a multi-object segmentation model from several public datasets with related masks, using self-learning and cross-dataset mask labeling. Afterward, we collaborate with hepatologists to correct lesion masks by referring to imaging and pathology reports. 
Second, exploring the confidence maps from the 3D U-Net models~\cite{3DUnet}, \cite{Isensee2020nnUNetAS}, we design multi-sensitivity algorithms to generate detection results from segmentation, maximizing information utilization. Third, we design the lesion-shuffle training scheme, which leads to more robust models for differentiating true lesions from hard negatives. Fourth, we conduct both the lesion-level and patient-level evaluations in the screening as well as the diagnosing setting, with several new practical metrics. 

\begin{figure*}[t]
	\centering
    \hfill
	\includegraphics[width=0.98\linewidth]{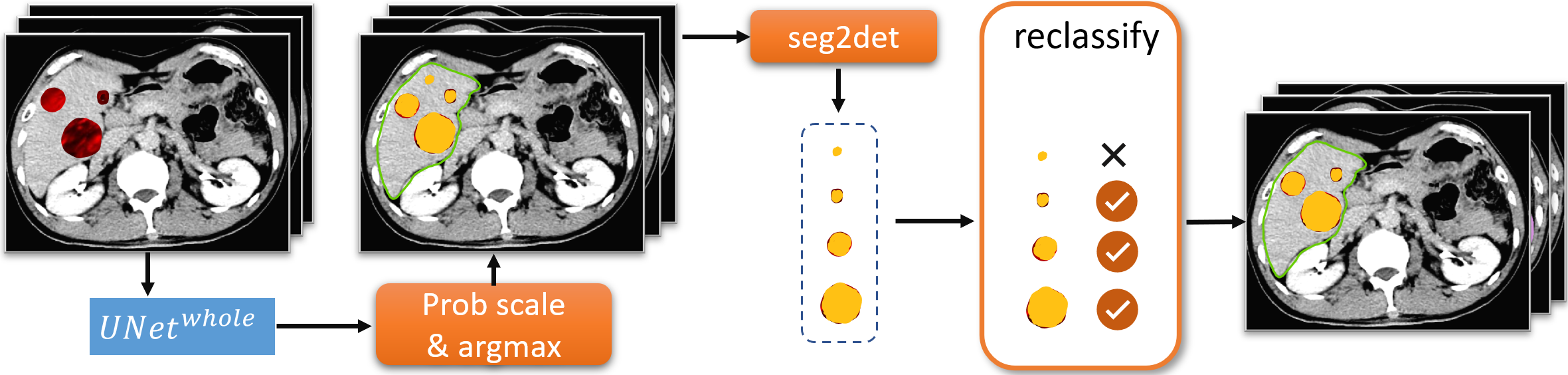}
	\hfill
	\caption{The proposed workflow. The 3D CT volume goes through the $UNet^{whole}$ model to get the 3D probability maps, and we adjust the lesion sensitivity to generate segmentation masks at different lesion sensitivities. We design the ``seg2det'' algorithm to generate precise lesion proposals and train a dedicated reclassification module to filter out false alarms.}
	\label{fig:system}
\end{figure*}

\section{Related work}
Several organizations have proposed guidelines for liver cancer diagnosis via CT scanning~\cite{HCCguidelineAsia2017}\cite{CTMRIHCCDetquality}\cite{HCCGuidelineConsensus}\cite{Roberts2018ImagingFT}\cite{HCCGuidelineEurope}. The machine models must follow these medical principles, in order to produce clinical-relevant results. Multi-phase contrast-enhanced CT and MRI form the cornerstone for the detection and diagnosis of hepatocellular carcinoma (HCC), especially for patients with cirrhosis~\cite{CTMRIHCCDetquality}\cite{HCCimaging}. The LI-RADS~\cite{LIRADS}, integrated into the HCC clinical practice guidelines by the AASLD~\cite{HCCguidelineByAASLD}, provides standardization for HCC imaging in the contexts of screening, surveillance, diagnosis, and treatment response assessment. 

Existing studies tackle challenges from various aspects~\cite{autoDetDelineationHCC}\cite{Jin2020RAUNetAH}, such as the organ contexts, lesion appearances, and data labeling. Many focus on the multiple-stage workflow~\cite{Bellver2017DetectionaidedLiverLesionSeg}\cite{PulmonaryNoduleDetFPReductionMultiView} to disentangle task complexity. Some works address the lesion ambiguity challenge by multiple view~\cite{PulmonaryNoduleDetFPReductionMultiView}, coarse to fine classification~\cite{DetClsFocalLiverLesion2021}, multi-phase CT images~\cite{DetClsFocalLiverLesion2021}. Some work addresses the data scarcity challenge by data augmentation, generative adversarial network (GAN)~\cite{LiverLesionGANSyntheticDA}, and semi-supervised learning~\cite{semiDeepTransferBenignMaligPulmonary}. 

The two-stage paradigm works well in many CT-based lesion detection tasks~\cite{CADeCADxLungCancer}\cite{AutoLungNoduleDet2stage}\cite{Han2017AutomaticLL}\cite{Christ2016AutomaticLA}\cite{Alirr2020DeepLA}\cite{Yasaka2018DeepLW}\cite{Azer2019DeepLW}\cite{Chlebus2018AutomaticLT}. As organs pack together closely in the abdomen, the boundaries can be vague due to the inter-organ appearance similarity. The disproportionate volumes between organs and lesions also impede the model training. One form of the two-stage pipeline~\cite{twoStageLiverDetTumorSeg} localizes the target organ first and then uses a dedicated model to learn finer-level lesion patterns. This coarse-to-fine approach helps reduce false positive lesions and better distinguish lesion types. 
In another coarse-to-fine pipeline ~\cite{Zhang2020DeepLILevvelSetLiverTumorSeg}, they employ the level-set method to further refine the tumor segmentation. The object-based post-processing~\cite{autoTumorSegWithObjectPost} can also be employed in the second stage.


Machine models detect lesions based on pixel-level or region-level probabilities, along with judging thresholds. However, a model optimized for high precision may overlook malignant lesions, while a model with high recall may produce an excessive number of false positives, leading to radiologists' extra ruling out time and even unnecessary follow-up tests ~\cite{CADeCADxLungCancer}. 
Therefore it is critical to make the detecting threshold reasonable. The HPVD model~\cite{HPVD} accepts any combination of contrast-phase inputs with adjustable sensitivity depending on the different clinical settings.

Although there have been several public CT datasets for liver tumor detection~\cite{Simpson2019ALA}\cite{Ma2022AbdomenCT1KIA}, many shortcomings and challenges exist. The medical segmentation decathlon (MSD)~\cite{Simpson2019ALA} collected data from several medical centers, containing abdominal CT images covering multiple organs (pancreas, lung, prostate, liver, colon). The AbdomenCT-1K~\cite{Ma2022AbdomenCT1KIA} reassembles data from existing datasets such as MSD and LiTS~\cite{LiTS}. However, these datasets are limited in terms of scale, CT phase availability, disease distribution, and lesion type labels. The clinical diagnosis process follows strict routines not addressed in current public datasets. Multi-phase CT images are required for accurate detection and diagnosis, which is also absent. 


Semi-supervised learning~\cite{mitSemiSupervised} can better utilize the knowledge in existing labeling to create new labeling~\cite{Huo2020HarvestingDA}\cite{learnFromMultiDatasets}\cite{semiDeepTransferBenignMaligPulmonary}. In the pseudo-labeling and self-training settings~\cite{surveyDeepSemiSupervised}, semi-supervised learning enhances the labeling gradually. 



\section{Methodology}
The proposed workflow (Fig.~\ref{fig:system}) consists of three steps: whole-CT segmentation, transforming multiple-sensitivity segmentation into detection, and lesion reclassification. For the first step, the $UNet^{whole}$ model segments 13 classes, including six major types of liver lesions (HCC, intrahepatic cholangiocarcinoma [ICC], metastasis [Meta], hemangioma [Hem], focal nodular hyperplasia + implants + others [Other], and Cyst) and seven organs\&vessels (liver, hepatic vessels, portal\&splenic vein, gallbladder, spleen, pancreas, and stomach) as in Fig.~\ref{fig:ctlabelerorgans}. The second and third steps 
maximize the segmenting information utilization to refine the detecting results.

From a holistic view, the system should output comprehensive information about individual lesions in the liver, including locations, 3D extractions (masks), volumes, densities, etc. The system output should also preserve suspicious and ambiguous information from the original CT scanning, which may serve as a clue for early disease alarm. From an implementation view, the system components would include voxel classification in the 3D image space, voxel clustering for bio-structures, and lesion extraction. 

Combining holistic goals and practical implementations, the initial voxel classification should extract reliable and detailed information, while the following steps should effectively refine and retain information. The final results should reflect the truthiness and comprehensiveness of the system detections. Throughout the workflow, related information is extracted and refined, incurring as less distortion as possible. From the information view, it is important to select implementing operations, intermediate representations, and evaluation metrics that preserve information and entropy.  


\begin{figure}[t]
	\centering
        \includegraphics[width=0.98\linewidth]{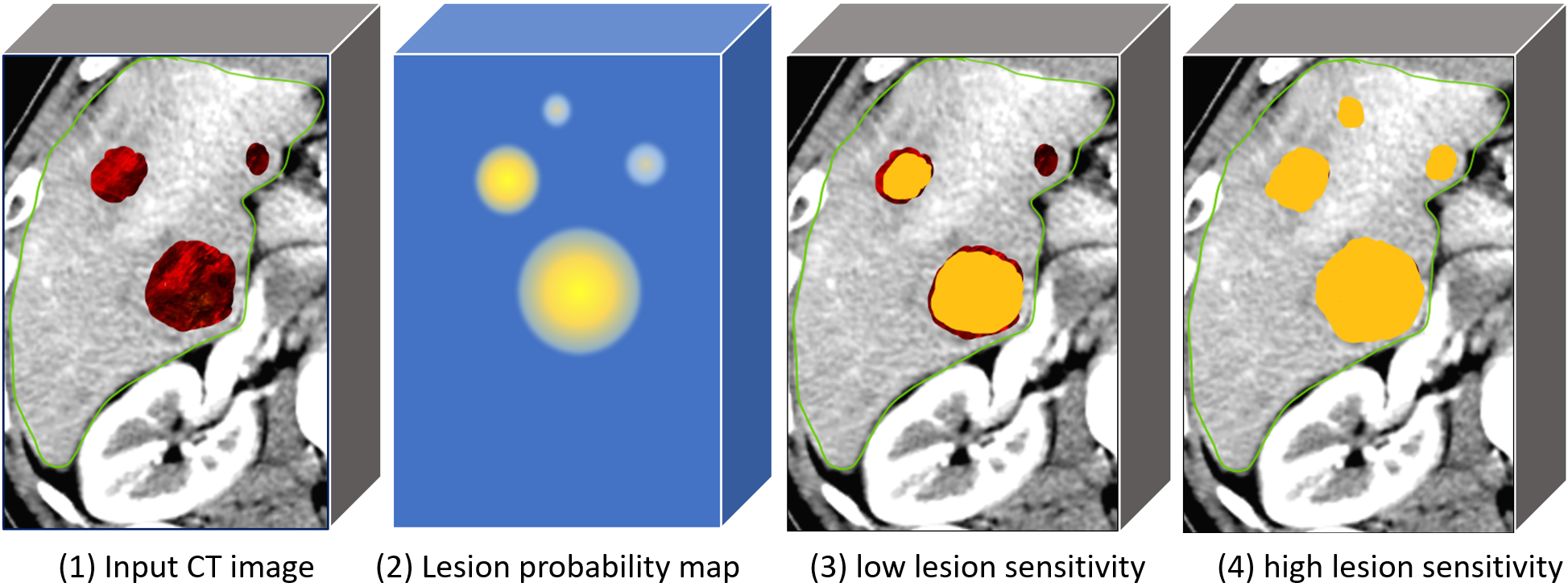}
	\caption{Illustration of lesion probability maps and segmentation results of varied sensitivities. The $UNet^{whole}$ generates separate confidence maps (2) for six lesion types. The small lesion representations in each 3D map are not stable. Decreasing or increasing the lesion sensitivity may lead to miss-detections (3) or false alarms (4). 
 }
	\label{fig:scale}
\end{figure}

\subsection{Multi-Sensitivity Segmentation}
\label{sec:method_multisens}
We employ the 3D nnUNet~\cite{Isensee2020nnUNetAS} as the $UNet^{whole}$ backbone, which takes in one or multiple 3D CT volumes ($\mathbf{I} \in \mathbb{R} ^{Z\times Y\times X}$, $\mathbf{I}$ is the volume) and generates probability confidence maps $\mathbf{Prob} \in \mathbb{R} ^{13\times Z\times Y\times X}$ for six liver lesion types and seven organs\&vessels,
\begin{equation}
\mathbf{Prob} \gets UNet^{whole} (\mathbf{I})
\label{eqn:nnunet}
\end{equation}

By default, the UNet applies the \textit{argmax} function on $\textbf{Prob}$ for generating the final 3D segmentation mask $\mathbf{ Mask} \in \mathbb{R} ^{Z\times Y\times X} $. Specifically, for each voxel $(z,y,x)$ in the 3D space, the \textit{argmax} compares all 13 individual probability maps to find the most confident one and assigns the corresponding label as the voxel value $\mathbf{Mask}^{pred}_{z,y,x}$ in the segmentation mask,  
\begin{equation}
\mathbf{Mask}^{pred}_{z,y,x} \gets \underset{i}{argmax} ( \mathbf{Prob_{i,z,y,x}}) ,\  \ i \in [1,13] 
\label{eqn:argmax}
\end{equation}

The lesions (label index: $1$--$6$) have much fewer volumes than organs ($7$--$13$), leading to imbalanced prediction tendencies in the probability maps. Treating all 13 labels equally during segmentation map generation results in miss-detection (i.e., false negative [FN]) of small or less obvious lesions, as shown in Fig~\ref{fig:scale} (3). By examining the individual confidence maps $\mathbf{Prob_i},\ i \in [1,13]$, we find that there is a wealth of explorable information pertaining to lesion detection. For example, in Fig.~\ref{fig:scale} (2), the lesion probability map has strong representations for large and middle tumors, but the small tumor has weak confidence. By setting $f=4$ in Equation~(\ref{eqn:probscale}) before applying the \textit{argmax} in Equation~(\ref{eqn:argmax}), the high lesion sensitivity segmentation (in Fig.~\ref{fig:scale} (4)) represents weak confidence signals for small lesions much better. The ``Prob scale'' process in Fig. \ref{fig:system} is defined as 
\begin{equation}
\mathbf{Prob}[i,:,:,:] \gets \mathbf{Prob}[i,:,:,:] \times \textit{f}\ ,\  \ i \in [1,6]
\label{eqn:probscale}
\end{equation}

We then develop the ``seg2det'' algorithm, which precisely separates individual lesions from the $\mathbf{Mask} \in \mathbb{R} ^{Z\times Y\times X}$. The lesions of one type all have the same label in $\mathbf{Mask}$, making it difficult to analyze individual lesions. There is no fixed shape for each lesion, and disentangling them requires analysis of the 3D boundary. The ``seg2det'' algorithm firstly separates all individual lesions in the 2D axial-view slices, then checks the lesion overlapping between neighboring slice pairs. If two lesions in neighboring slices have overlapping pixels in the 2D space, then they are considered from the same lesion in the 3D space. 
\begin{equation} 
\mathbf{Det}^{gt} \gets seg2det (\mathbf{Mask}^{gt})
\label{eqn:seg2detGT}
\end{equation}
\begin{equation} 
\mathbf{Det}^{pred}_{f=factor} \gets seg2det (\mathbf{Mask}^{pred}_{f=factor})
\label{eqn:seg2detPred}
\end{equation}
\begin{equation} 
\mathbf{TP}, \mathbf{FL}, \mathbf{FP}, \mathbf{FN} \gets match (\mathbf{Det}^{gt}, \mathbf{Det}^{pred})
\label{eqn:matchGt}
\end{equation}
\begin{equation} 
\mathbf{Det}^{pred}_{same}, \mathbf{Det}^{pred}_{diff} \gets match ( \mathbf{Det}^{pred}_{f=1}, \mathbf{Det}^{pred}_{f=4})
\label{eqn:matchPred}
\end{equation}

We apply the ``seg2det'' algorithm on both the ground truth $\mathbf{Mask}^{gt}$ and the predicted $\mathbf{Mask}^{pred}_{f=scale}$ to generate lesion detection sets $\mathbf{Det}^{gt}$ and $\mathbf{Det}^{pred}_{f=factor}$, each with precise 3D locations and masks. Further, we develop the ``match'' algorithm, which compares two lesion detection sets and outputs the overlapping relationship. There can be three matching results for each lesion site between $\mathbf{Det}^{gt}$ and $\mathbf{Det}^{pred}$, which are (1) overlap with the same label (true positive [TP]), (2) overlap with different labels (false label [FL]), and (3) lacking correspondence (false positive [FP] or false negative [FN]). 

\subsection{Lesion Reclassification}
\label{sec:reclassify}

\begin{figure}[bt!]
	\centering
	\includegraphics[width=0.98\linewidth]{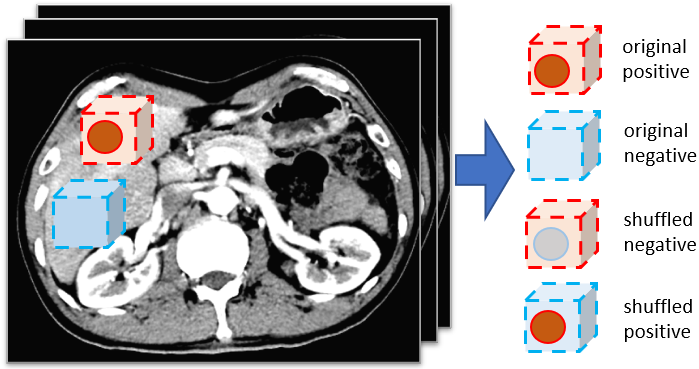}
	\caption{The lesion shuffling process is based on the lesion mask and detection results. The patches are randomly sampled around the lesion or in \textit{lesion-free} areas in the liver. The shuffling exchanges a lesion with normal liver textures, creating augmentations for training or inferencing. }
	\label{fig:shuffle}
\end{figure}

The drawback of high lesion sensitivity segmentation (Fig~\ref{fig:scale} (4)) is an increased occurrence of FPs on suspicious liver nodules or normal textures. To enhance the contrast between normal liver textures and TP/FP lesions in the predicted lesion detection set $\mathbf{Det}^{pred}$, we employ a dedicated segmentation model (based on nnUNet~\cite{Isensee2020nnUNetAS}), denoted as $UNet^{patch}$, which is trained to classify voxels within liver-lesion mixture patches. All the voxels outside the liver are removed (mask out). During inference, each lesion has multiple randomly augmented patches and the lesion truthiness is measured by the average classification score.

\begin{equation} 
\mathbf{Det}^{ReCls} \gets reclassify ( \mathbf{Det}^{pred})
\label{eqn:reclassify}
\end{equation}

We design the ``shuffle'' algorithm (Equation~(\ref{eqn:shuffle})) which creates 3D augmentation patches of the individual lesions. With the exact lesion mask and liver mask in the 3D space, we are able to crop patches in the lesion-free liver area and then randomly transplant the lesion into patches, as shown in Fig~\ref{fig:shuffle}. This context-aware augmentation process effectively utilizes possible lesion surroundings within the same patient's liver. By additionally shuffling $\mathbf{Det}^{pred}_{FP}$ and $\mathbf{Det}^{pred}_{FN}$, the generated patches serve as hard cases for training augmentation. 
\begin{equation} 
\mathbf{I}_{patch}, \mathbf{Mask}_{patch} \gets shuffle ( \mathbf{Det}, \mathbf{I})
\label{eqn:shuffle}
\end{equation}

\begin{figure}[bt!]
	\centering
	\includegraphics[width=0.98\linewidth]{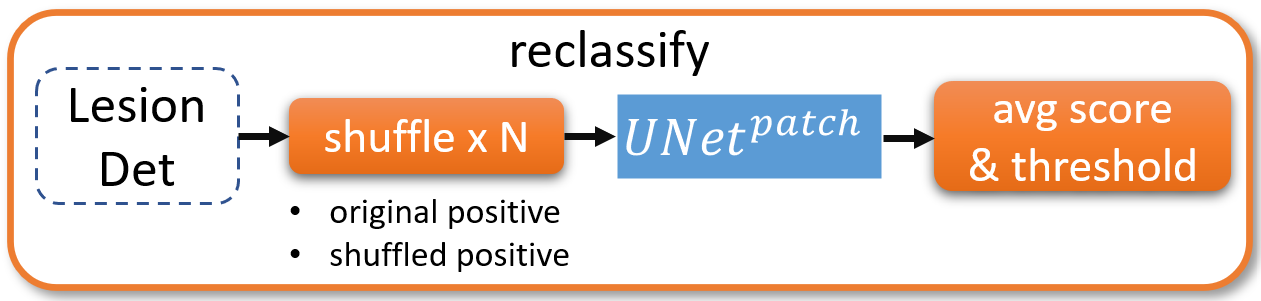}
	\caption{The reclassification process. Each lesion proposal is shuffled within the liver, to create $N$ augmented patches for the $UNet^{patch}$ to segment. The average score is compared with a threshold to re-classify this lesion.}
	\label{fig:reclassify}
\end{figure}

During inference, the ``shuffle'' algorithm generates $N$ original positive and shuffled positive patches. The $UNet^{patch}$ predicts on each patch $\mathbf{I}_{patch,n}$ to generate $\mathbf{Mask}_{patch,n}^{pred}$ in Equation~(\ref{eqn:unetpatch}). The newly predicted $\mathbf{Mask}_{patch,n}^{pred}$ and the originally predicted $\mathbf{Mask}_{patch,n}$ are binarized and compared for lesion overlapping in Equation~(\ref{eqn:avgpatch}). And the overlapping ratio in all $N$ patch comparisons serves as the average score for the lesion. If the score is below a threshold, the lesion will be discarded in the $\mathbf{Det}^{ReCls}$. As a result, the FPs especially in high sensitivity segmentation are effectively reduced. Such lesion reclassification process is shown in Fig~\ref{fig:reclassify}.

\begin{equation} 
\mathbf{Mask}^{pred}_{patch,n} \gets argmax(UNet^{patch} ( \mathbf{I}_{patch,n}))
\label{eqn:unetpatch}
\end{equation}

\begin{equation} 
score \gets \frac{1}{N} \Sigma^{N}_{n=1} ( bin(\mathbf{Mask}^{pred}_{patch,n}, \mathbf{Mask}_{patch,n}))
\label{eqn:avgpatch}
\end{equation}

\section{Experimental Methods}
\label{sec:Experiments}
\subsection{Data and Annotation}
\subsubsection{Data collection}
Our dataset consists of patients collected between 2008 and 2018 at Chang Gung Memorial Hospital, Taiwan, ROC. We follow the Helsinki declaration with ethical permission number IRB-201800187B0 (liver tumor detection through CT images). The received CT images are in NIfTI format with patient information removed to protect privacy. The CT scanners include GE, SIEMENS, and TOSHIBA. The in-plane resolution ranges from 0.6 mm $\times$ 0.6 mm to 1.0 mm $\times$ 1.0 mm, and the slice thickness is 5.0 mm. Each axial slice size is 512 $\times$ 512, and the slice number varies from 35 to 78. The four-phase (NC, AP, VP, and DP) CT images are aligned by the DEEDS~\cite{DEEDS} algorithm. The malignant lesions are manually annotated with 3D bounding boxes by an expert physician (C-T. C) with referring to the radiology and pathology reports~\cite{HPVD}.

\begin{figure}[t]
	\centering
    \hfill
	\includegraphics[width=0.98\linewidth]{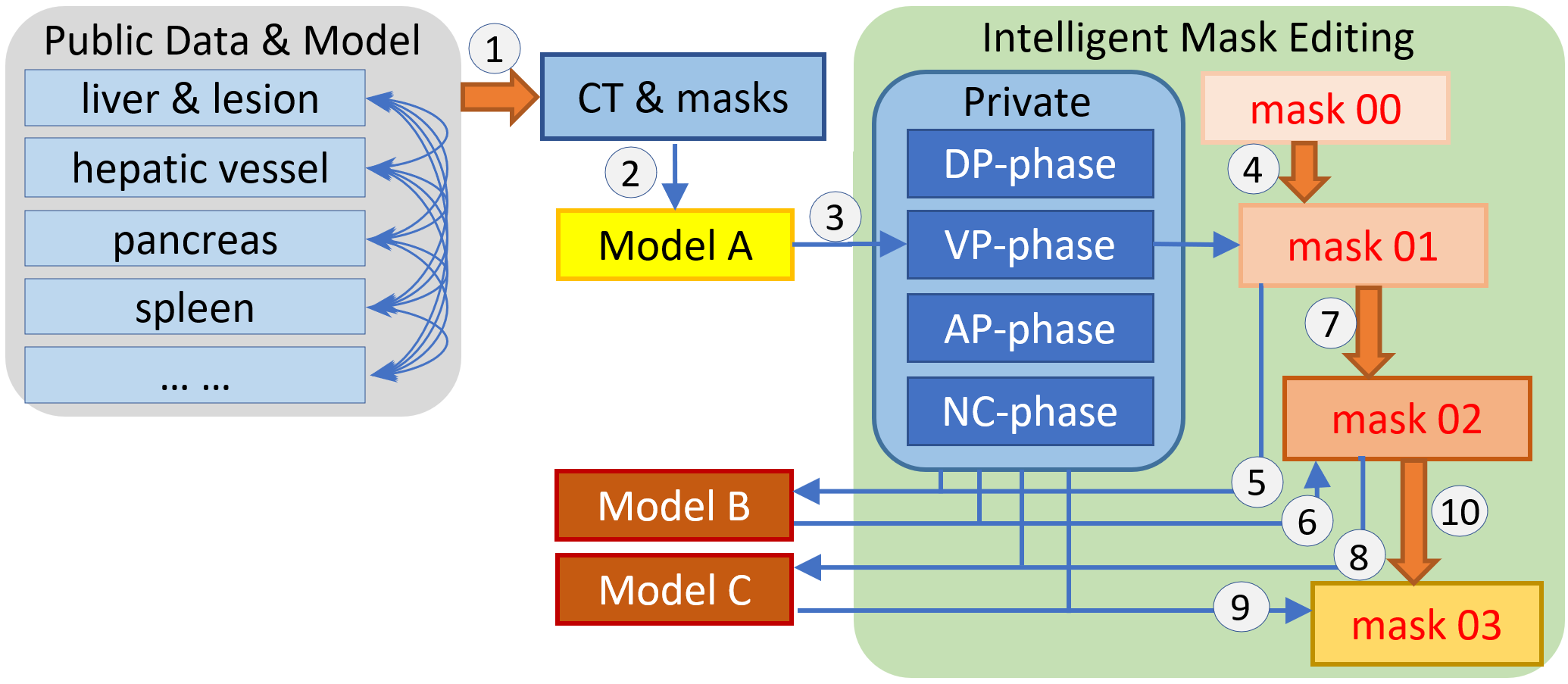}
	\hfill
	\caption{Data curation steps. The public datasets only have Venous-Phase CTs, and they do not differentiate lesion types. The \textit{Intelligent Editing} process utilizes the \textit{CT Labeler} in Fig~\ref{fig:ctlabelerorgans}, which automatically compares lesion masks for efficient updating in Steps $4,7,10$. Train/inference steps $5,6,8,9$ focus on segmenting six lesion types and seven organ types in the four-phase dataset.}
	\label{fig:curationsteps}
\end{figure}

\subsubsection{Curation steps}
In the curation steps (Fig~\ref{fig:curationsteps}), we apply the semi-supervised learning on the public datasets (step \textcircled{1}) and on our private dataset (steps \textcircled{5}, \textcircled{6}, \textcircled{8}). The public CTs and masks are from seven segmentation tasks in the MSD challenge~\cite{MSD}. The cross-inference on public datasets using pre-trained nnUNet models~\cite{Isensee2020nnUNetAS} produces 900 CTs with seven organs\&vessels masks and liver lesion masks. The labeling knowledge (masks) is transferred to our dataset by Model A inferencing on VP-phase to get mask 01 in step \textcircled{3}. 

Our CT Labeler's convenient functions, such as automatic information (image, mask, reports) loading, automatic windowing, 3D connected-region computation, lesion boundary visualization, and mask editing (updating/copying/deleting), substantially increase labeling efficiency. In Fig~\ref{fig:curationsteps}, mask 00 are bounding boxes containing pathologically-confirmed malignant tumors, and two physicians (C-T. C and C-W. P) update mask 01 for lesions and organs in step \textcircled{4}. Existing and newly discovered lesions are re-examined against the radiology and pathology reports, divided into six sub-types. Trained on four-phase data with corrected mask 01, Model B is much more reliable than Model A and produces mask 02. Then we do another round of mask updating and model retraining. The scale and quality of masks get stably improved from mask 01 to mask 03. In the end, we get 1,631 cases, with high-quality 3D masks for seven organs\&vessels and six liver lesion types. 

\subsubsection{Data split and distribution}
We divide the 1,631 cases into a train-validation set (n=1,300) and a test set (n=331). We employ the default 5-fold training-validation split and model-ensemble prediction in nnUNet~\cite{Isensee2020nnUNetAS}. The lesion-level statistics of the test set are in Table~\ref{tab:statisticsTesting}. The lesion has varied sizes/volumes, and the majority of non-HCC lesions are smaller than 16 cm$^3$. If multiple lesion types are present in one patient, the patient-level main lesion (\ie patient class [PatCls]) is defined as the highest-priority lesion type or ``normal''. The lesion priority order is set as HCC $>$ ICC $>$ Meta $>$ Hem $>$ Other $>$ Cyst, considering the malignancy degree. The ``classifyPatient'' function in Equation~(\ref{eqn:classifyPatient}) implements the above lesion classification rules.

\vspace{2mm}
\begin{table}[h]
	\caption{The numbers of lesions stratified by volume distribution and lesion types in the test set. Lesions smaller than 0.5 $cm^3$ are excluded.}
	\centering
	
\scriptsize
\begin{tabular}{|c|c|c|c|c|c|c|c|}
\hline
Type \textbackslash \textbf{V} $cm^3$  & 0.5$\sim$2 & 2$\sim$4 &4$\sim$8& 8$\sim$16 & 16$\sim$64 & $>$64 & \textbf{Total}  \\\hline
HCC    & 22         & 28       & 36       & 47        & 60         & 66               & 259    \\\hline
ICC  & 0          & 0        & 1        & 2         & 7          & 3                & 13      \\\hline
Meta   & 69         & 27       & 32       & 18        & 24         & 22               & 192     \\\hline
Hem  & 1          & 2        & 1        & 0         & 3          & 5                & 12       \\\hline
Other  & 25         & 8        & 7        & 11        & 8          & 2                & 61      \\\hline
Cyst   & 51         & 7        & 8        & 1         & 2          & 0                & 69      \\\hline
\textbf{All}    & 168        & 72       & 85       & 79        & 104        & 98               & 606   \\\hline
\textbf{Malig}  & 91         & 55       & 69       & 67        & 91         & 91               & 464    \\\hline
\end{tabular} \label{tab:statisticsTesting}
\end{table}

\begin{equation}
\textit{PatCls} \gets \textit{classifyPatient}( \mathbf{Dets})
\label{eqn:classifyPatient}
\end{equation}


\subsection{Implementation}


\subsubsection{Configurations for lesion shuffling}
In lesion reclassification (Sec. \ref{sec:reclassify}), we fix the patch size (\ie input croppings) to be $16\times128\times128$ voxels (approximately 8 $\times$ 10 $\times$ 10cm), enough for filtering out FP lesions. Lesions with a volume larger than  64cm$^3$ in $Det^{pred}$ will not go through the lesion reclassification process. 

In $UNet^{patch}$ training, to generate as many realistic input patches as possible, we adopt four shuffling schemes in Fig.~\ref{fig:shuffle}. The original positive patches are randomly cropped around the lesion, 
and the original negative patches are randomly sampled without containing any lesion. 
The shuffled positive patch transplants a lesion into the original negative patch, and the shuffled negative patch replaces the lesion in the original positive patch with normal liver texture. The shuffling process generates both CT images and ground-truth masks for individual patches. Each lesion has around $20$ augmentation patches under any shuffling scheme during $UNet^{patch}$ training.

In the reclassification inference process (Fig.~\ref{fig:reclassify}), we generate $N=10$ augmented lesion-containing patches for each detected lesion. The binarize function in Equation~(\ref{eqn:avgpatch}) calculates the overlapping volume between $UNet^{whole}$-generated $\mathbf{Mask}_{patch,n}$ and the $UNet^{patch}$-generated $\mathbf{Mask}^{pred}_{patch,n}$. If the average score is smaller than a threshold of 0.5cm$^3$, then this lesion is discarded (\ie reclassified as FP).

\subsubsection{Multi-sensitivity detection}
We set $f=1.0$ and $f=4.0$ (Equation (\ref{eqn:probscale})) as lesion probability factors for low sensitivity and high lesion sensitivity. 
As a result, the corresponding detection sets after the reclassify module are $Det^{ReCls}_{f=1}$ and $Det^{ReCls}_{f=4}$, respectively. We then apply the ``match'' in Equation~(\ref{eqn:matchPred}) to generate the common set $Det^{ReCls}_{same}$ and difference set $Det^{ReCls}_{diff}$. The lesions proposals in $Det^{ReCls}_{same}$ are more reliable, while those in $Det^{ReCls}_{diff}$ are mostly from $Det^{ReCls}_{f=4}$ only.

\vspace{2mm}
\subsection{Experiment Settings} 
\subsubsection{Evaluation metrics}
We measure the system performance at the lesion level and patient level. The lesion-level detection results enable measuring the TP, FL, FP, and FN (defined in Section~\ref{sec:method_multisens}) for all test cases, and we can derive the overall lesion detection recall and precision, \ie $recall = \frac{TP}{TP+FN}$, and $prec = \frac{TP}{TP+FL+FP}$. We calculate these metrics for each lesion type. 
At the patient-level, we use the classification accuracy of the patient-level main lesion as the metric, \ie $acc = \frac{1}{N} \sum^{N}_{i} (PatCls^{pred}_{i} == PatCls^{gt}_{i})$ ($PatCls$ is defined in Equation~(\ref{eqn:classifyPatient})).


\begin{figure}[t]
	\centering
    \hfill
	\includegraphics[width=0.98\linewidth]{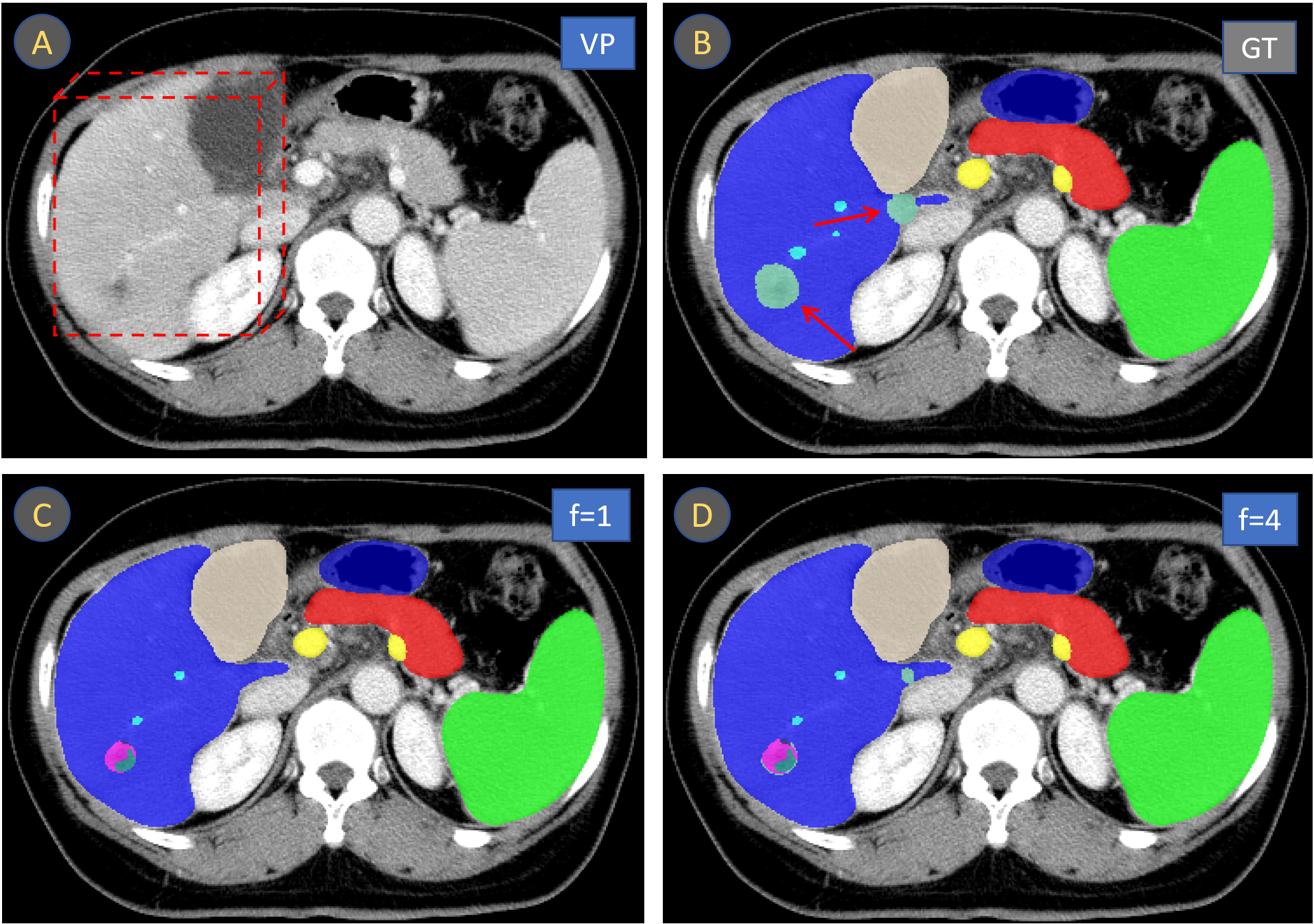}
	\hfill
	\caption{Case 01 masks of different lesion sensitivities. (A) The selected axial slice is from the venous-phase CT scanning. 
 (B) Ground-truth mask overlay on the CT image slice, with the arrows pointing to two HCC tumors. (C) The automatically predicted mask by using lesion sensitivity $f=1$ on $UNet^{whole}$ probability maps, resulting in one miss detection (FN) and one False Labeling (FL); the Base model derives $Det^{pred}_{f=1}$ directly from this mask. (D) The automatically predicted mask by using lesion sensitivity $f=4$, detecting both HCC tumors; the High+ReCls model derives $Det^{pred}_{f=4}$ from this mask.}
	\label{fig:workflowWholeSegs}
\end{figure}

\begin{figure}[t]
	\centering
    \hfill
	\includegraphics[width=0.98\linewidth]{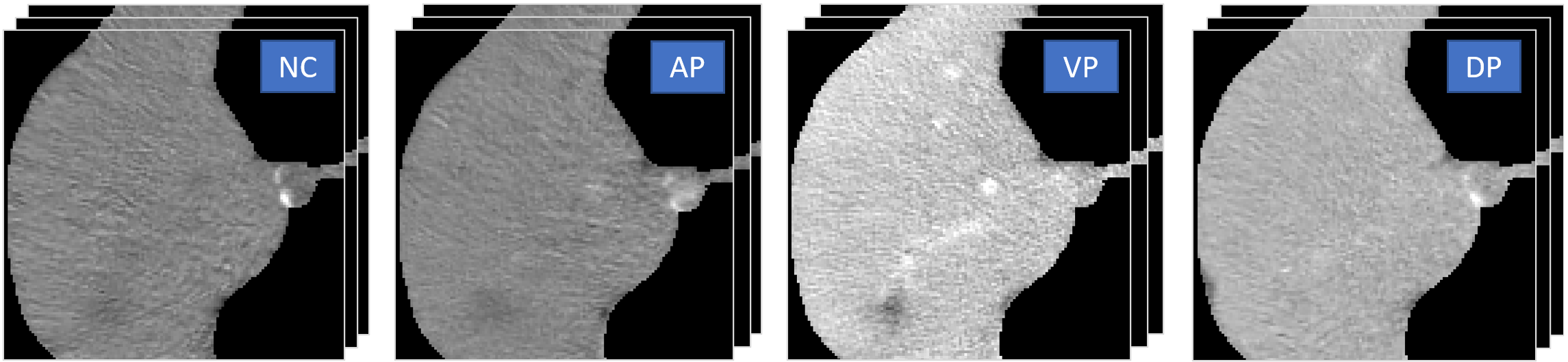}
	\hfill
	\caption{The patch cropping ($16\times128\times128$) in Fig.~\ref{fig:workflowWholeSegs} (A) (dashed cube), containing HCC tumors while excluding other organs. The $UNet^{patch}$ model in reclassification module would train and test on these curated patches, and some are augmented by the lesion shuffle process. }
	\label{fig:workflowPatch}
\end{figure}

\begin{figure}[t]
	\centering
    \hfill
	\includegraphics[width=0.98\linewidth]{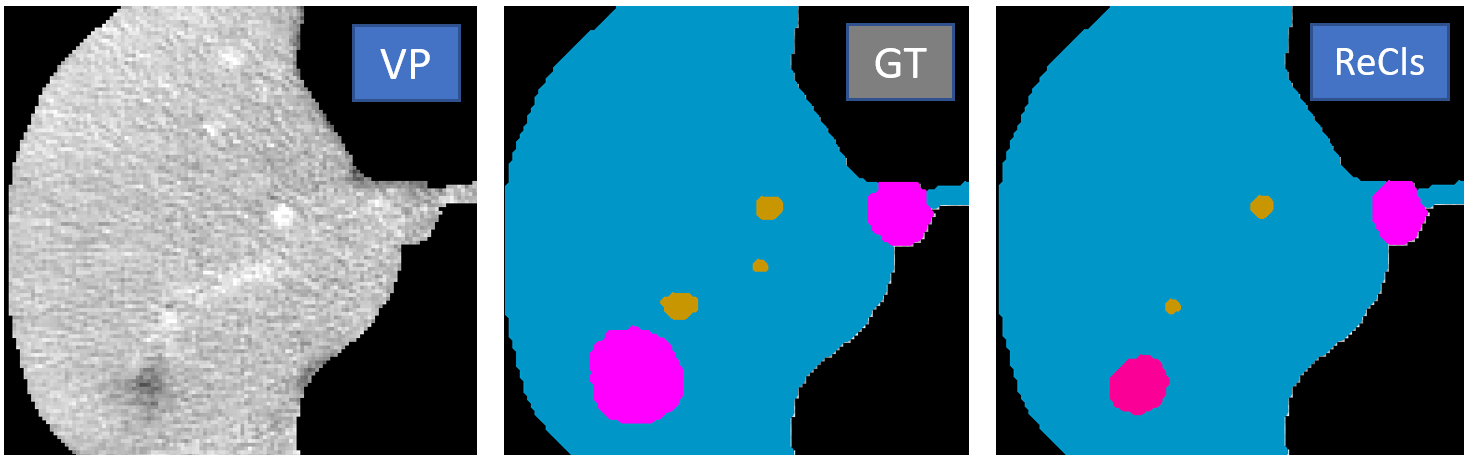}
	\hfill
	\caption{ Reclassified result for Fig.~\ref{fig:workflowPatch}. The two ground-truth HCC are in pink. The $UNet^{patch}$ segmentation mask (ReCls) incorrectly identifies one HCC as Meta (red) due to the absence of typical HCC characteristics described in LI-RADS.}
	\label{fig:workflowPatchResult}
\end{figure}

\subsubsection{System workflow}

We illustrate the three-step workflow with intermediate results in Figs.~\ref{fig:workflowWholeSegs}--\ref{fig:workflowPatchResult}. In Fig.~\ref{fig:workflowWholeSegs}, Case 01 has two HCC tumors, one is small and located at the liver boundary, and the other lacks typical HCC characteristics. These challenges are seen in Fig.~\ref{fig:workflowWholeSegs}~(C), where the smaller one is missed and the larger one has confusing predictions. The proposed High+ReCls (high sensitivity + reclassification) model relies on high lesion sensitivity detections in Fig.~\ref{fig:workflowWholeSegs}~(D) and re-classifies crops in Fig.~\ref{fig:workflowPatch}, to produce final ReCls results in Fig.~\ref{fig:workflowPatchResult}. 

\subsubsection{Model variations}
For the baseline (Base), we adopt the default nnUNet segmentation mask in the first step, which uses $f=1$ in Equation~(\ref{eqn:probscale}). Base does not have the reclassification module (the third step). So the Base detection results would have many FNs due to the low lesion sensitivity in the first step, as well as many FPs due to the lack of lesion filtering in the third step.

In the Base+ReCls model, we add the reclassification module. It leads to reduced FPs correspondingly. In the High+ReCls model, we set lesion sensitivity $f=4$ in Equation~(\ref{eqn:probscale}) for the segmentation mask generation (first step), which leads to more detections in the second step. Therefore the High+ReCls model has both reduced FNs and reduced FPs. Compared to Base+ReCls, High+ReCls would have fewer FNs at the cost of a few more FPs. Most of the time, these two models output the same PatCls for each patient case through Equation~(\ref{eqn:classifyPatient}). When a patient-level difference does exist, it is usually originated from suspicious lesions, and worth human verification. 

In the Base+ReCls w/o LS model, we remove lesion shuffling (LS) in the reclassify module to investigate its effect. In this setting, the training and testing of $UNet^{patch}$ only involve randomly cropped patches around the lesions, and there is no shuffled positive or shuffled negative augmentation.

In the diagnosis setting, the contrast-enhanced CT scanning images ($I^{NC,AP,VP,DP}$) are aligned before being fed into the $UNet^{whole}$. While in the screening setting, there is only the noncontrast CT image ($I^{NC}$). We run experiments of Base, Base+ReCls, Base+ReCls w/o LS, and High+ReCls in both settings under the same experimental protocols.

\section{Results}
\label{sec:Results}
\subsection{Lesion-Level Performances} 
Table~\ref{tab:lesion_level_diagnosis} and Table~\ref{tab:lesion_level_screen} summarize the lesion-level performance of different models on the testing set. The Malig group includes HCC, ICC, and Meta. 
It is worth mentioning that our dataset contains mostly patients with liver diseases and the performance of the screening setting may vary in real practice. In Table~\ref{tab:statisticsTesting}, there are $606$ lesions in total and $464$ of them are malignant. The majority of patients with malignant lesions have large lesions (volume $>$ 4cm$^3$). The benign lesions (Hem, Other, Cyst) have less representation and are usually of small size (volume $<$ 8cm$^3$). In real-world screening scenarios, most patients are free of malignant tumors, with benign lesions outnumbering malignant ones and most lesions being small.

\begin{table*}[t]
	\caption{Lesion-level performance of different models in the diagnosis setting using four-phase CT (I$^{NC,AP,VP,DP}$, 331 testing cases). Using the ``seg2det'' and ``match'' algorithms, we precisely compare all individual lesions between the prediction and the ground truth. FN and FP are the numbers of false negative and false positive lesions, respectively, for each lesion type. }
	\centering
	\hfill
	\scriptsize
\begin{tabular}{|c|c|c|c|c|c|c|c|c|c|c|c|c|c|c|c|c|}
\hline
\multirow{2}{*}{Type \textbackslash Model} & \multicolumn{4}{c|}{Base} & \multicolumn{4}{c|}{Base+ReCls w/o LS} & \multicolumn{4}{c|}{Base+ReCls} & \multicolumn{4}{c|}{High+ReCls} \\\cline{2-17}
                                          & FN    & FP    & Precision     & Recall     & FN        & FP        & Precision         & Recall        & FN    & FP    & Precision     & Recall    & FN    & FP    & Precision     & Recall    \\\hline
HCC                                       & 16    & 7     & 84.1\%    & 88.4\%    & 26        & 1         & 89.4\%        & 84.9\%       & 17    & 1     & 87.0\%    & 88.0\%   & 12    & 8     & 84.6\%    & 89.9\%   \\\hline
ICC                                     & 0     & 3     & 50.0\%    & 61.5\%    & 0         & 1         & 57.1\%        & 61.5\%       & 0     & 1     & 57.1\%    & 61.5\%   & 0     & 1     & 57.1\%    & 61.5\%   \\\hline
Meta                                      & 24    & 8     & 85.9\%    & 77.5\%    & 40        & 5         & 86.2\%        & 68.8\%       & 24    & 7     & 86.5\%    & 77.5\%   & 13    & 8     & 85.5\%    & 82.3\%   \\\hline
Hem                                     & 0     & 2     & 60.0\%    & 46.2\%    & 2         & 0         & 60.0\%        & 50.0\%       & 1     & 0     & 66.7\%    & 50.0\%   & 1     & 1     & 54.5\%    & 50.0\%   \\\hline
Other                                     & 6     & 14    & 52.6\%    & 49.2\%    & 19        & 2         & 66.7\%        & 36.7\%       & 11    & 3     & 67.6\%    & 41.7\%   & 9     & 8     & 62.5\%    & 41.7\%   \\\hline
Cyst                                      & 0     & 5     & 87.5\%    & 88.9\%    & 7         & 1         & 94.4\%        & 82.3\%       & 3     & 2     & 91.8\%    & 88.9\%   & 5     & 4     & 88.9\%    & 88.9\%   \\\hline
\textbf{All}                                       & 46    & 39    & 80.5\%    & 79.5\%    & 94        & 10        & 86.0\%        & 73.5\%       & 56    & 14    & 84.9\%    & 78.8\%   & 40    & 30    & 82.5\%    & 81.0\%   \\\hline
\textbf{Malig}                                     & 40    & 18    & 95.6\%    & 88.7\%    & 66        & 7         & 98.1\%        & 83.1\%       & 41    & 9     & 97.8\%    & 88.5\%   & 25    & 17    & 95.9\%    & 91.8\%  \\\hline

\end{tabular} 
	\hfill
	\label{tab:lesion_level_diagnosis}
\end{table*}

\begin{table*}[t]
	\caption{Lesion-level performance of different models in the screening setting using only noncontrast CT (I$^{NC}$, 331 testing cases). }
	\centering
	\hfill
	\scriptsize
\begin{tabular}{|c|c|c|c|c|c|c|c|c|c|c|c|c|c|c|c|c|}
\hline
\multirow{2}{*}{Type \textbackslash Model} & \multicolumn{4}{c|}{Base} & \multicolumn{4}{c|}{Base+ReCls w/o LS} & \multicolumn{4}{c|}{Base+ReCls} & \multicolumn{4}{c|}{High+ReCls} \\\cline{2-17}
                                          & FN    & FP    & Precision     & Recall     & FN        & FP        & Precision         & Recall        & FN    & FP    & Precision     & Recall    & FN    & FP    & Precision     & Recall    \\\hline
HCC                                       & 30    & 27    & 72.5\%     & 83.9\%   & 49         & 2        & 79.8\%        & 77.3\%       & 34    & 6     & 77.6\%    & 82.7\%   & 24    & 15    & 73.6\%    & 86.7\%   \\\hline
ICC                                     & 0     & 2     & 44.4\%     & 30.8\%   & 0          & 1        & 44.4\%        & 30.8\%       & 0     & 1     & 44.4\%    & 30.8\%   & 0     & 1     & 44.4\%    & 30.8\%   \\\hline
Meta                                      & 39    & 7     & 88.5\%     & 60.0\%   & 63         & 2        & 93.8\%        & 50.0\%       & 39    & 3     & 93.1\%    & 60.0\%   & 26    & 10    & 85.1\%    & 62.6\%   \\\hline
Hem                                     & 2     & 0     & 100\%    & 33.3\%   & 3          & 0        & 100\%       & 33.3\%       & 2     & 0     & 100\%   & 33.3\%   & 1     & 0     & 100\%   & 33.3\%   \\\hline
Other                                     & 12    & 8     & 59.0\%     & 39.7\%   & 21         & 1        & 73.9\%        & 29.8\%       & 15    & 2     & 69.0\%    & 34.5\%   & 12    & 4     & 70.0\%    & 35.0\%   \\\hline
Cyst                                      & 2     & 3     & 81.6\%     & 95.4\%   & 5          & 3        & 82.1\%        & 90.2\%       & 3     & 3     & 81.3\%    & 93.8\%   & 3     & 5     & 83.3\%    & 93.8\%   \\\hline
\textbf{All}                                       & 85    & 47    & 76.1\%     & 71.2\%   & 141        & 9        & 82.3\%        & 63.5\%       & 93    & 15    & 80.8\%    & 70.0\%   & 66    & 35    & 77.0\%    & 72.6\%   \\\hline
\textbf{Malig}                                     & 69    & 36    & 91.0\%     & 80.8\%   & 112        & 5        & 98.5\%        & 72.3\%       & 73    & 10    & 97.3\%    & 80.1\%   & 50    & 26    & 93.2\%    & 85.3\%  \\\hline
\end{tabular} 
	\hfill
	\label{tab:lesion_level_screen}
\end{table*}

\subsubsection{Base model}
In the Base model, the nnUNet segmentation output ($\mathbf{Mask}^{pred}$, lesion sensitivity $f=1$) is fed directly to the ``seg2det'' and ``match'' steps to get the $Det^{pred}_{f=1}$, TP, FL, FP, FN results. In Table~\ref{tab:lesion_level_diagnosis}, the Base model has low FN and FP for the Cyst, and the recall reaches 88.9\%. However, the malignant types have high FN (40) and high FP (18), which we want to decrease. The precision and recall for ICC, Hem, Other are relatively low, due to unbalanced data representation and ambiguous lesion appearances. In the screening setting of Table~\ref{tab:lesion_level_screen}, FN and FP show substantial increases (85, 47). The precision and recall decrease substantially for all lesion types except the Cyst (81.6\%, 95.4\%). Many lesions lack the characteristic appearance under noncontrast CT scanning (I$^{NC}$) and are misclassified as Cyst. Compared with Ref.~\cite{Huo2020HarvestingDA} on the same data source (diagnosis setting) using a detection approach, our Base model achieves a 10\% higher F-score (86.2\% vs. 76.3\%) for HCC, effectively supporting our semantic segmentation-for-detection approach.

\subsubsection{Base+ReCls model}
For the Base+ReCls model results in Table~\ref{tab:lesion_level_diagnosis}, the FPs and precisions for all lesion types are improved substantially compared to the Base model, due to the re-classification module. The cost is the slight increase of FNs. In Table~\ref{tab:lesion_level_screen}, similar effects are observed. And the FP and precision for malignant improve even more (72\% reduction and 6.3\% increase).  

To investigate the influence of the Lesion Shuffling (LS) in the re-classification module, we conduct experiments of Base+ReCls w/o LS in Table~\ref{tab:lesion_level_diagnosis} and Table~\ref{tab:lesion_level_screen}. The ReCls w/o LS could still effectively filter out FPs for all lesion types, but it suppresses many true lesions leading to a large increase of FNs. Without the context-aware LS during augmentation, the $UNet^{patch}$ model in the re-classification module could not distinguish lesions from normal liver textures very well. 

\subsubsection{High+ReCls model}
The High+ReCls model sets lesion sensitivity $f=4$ in Equation~(\ref{eqn:probscale}) before generating the $Mask^{pred}_{f=4}$ in Equation~(\ref{eqn:argmax}) and $Det^{pred}_{f=4}$ in Equation~(\ref{eqn:seg2detPred}). Comparing High+ReCls with Base+ReCls in Table~\ref{tab:lesion_level_diagnosis}, the high lesion sensitivity segmentation can effectively reduce FNs, especially for malignant lesions, at the cost of increasing FPs. The recall rates do not catch up comparably due to false label (lesion type prediction error) detections among the improved portion. The high lesion sensitivity segmentation shows a similar influence for the screening setting in Table~\ref{tab:lesion_level_screen}.

\subsection{Patient-Level Performances} 
In the screening setting, the patient-level result may help with patient triage. 
In the diagnosis setting, when there are multiple lesions, the most severe one draws attention. In both settings/scenes, whether the patient has a malignant lesion or not is the primary question to answer. 
Therefore we compare the patient-level malignancy classification results of different models on the testing set in Table~\ref{tab:patient_level_malig}.

\begin{table}[t]
	\caption{Patient-level malignancy classification performance for different models on 331 testing cases. In both diagnosis and screening settings, sensitivity and specificity are measured for each model. The results of Base+ReCls \& High+ReCls only count in 319 cases with common PatCls predictions.}
	\centering
	\hfill
	\scriptsize

\begin{tabular}{|c|c|c||c|c|}
\hline
\multirow{2}{*}{Model \textbackslash Scene Metrics}  & \multicolumn{2}{c||}{Diagnosis of \textbf{Malig}}  &  \multicolumn{2}{c|}{Screening of \textbf{Malig}}  \\\cline{2-5}
                                                   & Sensitivity   & Specificity  & Sensitivity   & Specificity   \\\hline
Base                         & 97.7\% & 92.9\% & 95.0\% & 82.9\%  \\\hline
Base+ReCls w/o LS              & 96.2\% & 98.6\% & 89.3\% & 98.6\%  \\\hline
Base+ReCls                         & 97.3\% & 98.6\% &  92.3\% & 95.7\% \\\hline
High+ReCls                         & 99.2\% & 97.1\%  & 97.3\% & 95.7\% \\\hline
Base+Recls \& High+ReCls        & 99.2\% & 98.5\%   & 97.5\% & 98.4\%  \\\hline

\end{tabular}

	\hfill
	\label{tab:patient_level_malig}
\end{table}

We measure the sensitivity and specificity for malignancy classification in Table~\ref{tab:patient_level_malig} for both settings. In the diagnosis setting, the specificity increases by 5\% after adding the re-classification module to the Base model. The sensitivity of the High+ReCls model reaches 99.2\%, which is nearly 2\% higher than models with a base lesion sensitivity. These changes are more obvious in the screening setting, increasing by 12.8\% and 5\%, respectively.

There is a complementary relationship between Base+ReCls and High+ReCls in their predictions. From Table~\ref{tab:lesion_level_diagnosis} and Table~\ref{tab:lesion_level_screen}, the Base+ReCls has the advantage of lower FPs, while the High+ReCls has the advantage of lower FNs. Combining Base+ReCls \& High+ReCls could take both advantages. The Base+ReCls \& High+ReCls performance only count in patient cases where the two models output the same patient-level classification (PatCls), through the ``classifyPatient'' function in Equation~(\ref{eqn:classifyPatient}).

\begin{figure}[t]
	\centering
    \hfill
	\includegraphics[width=0.98\linewidth]{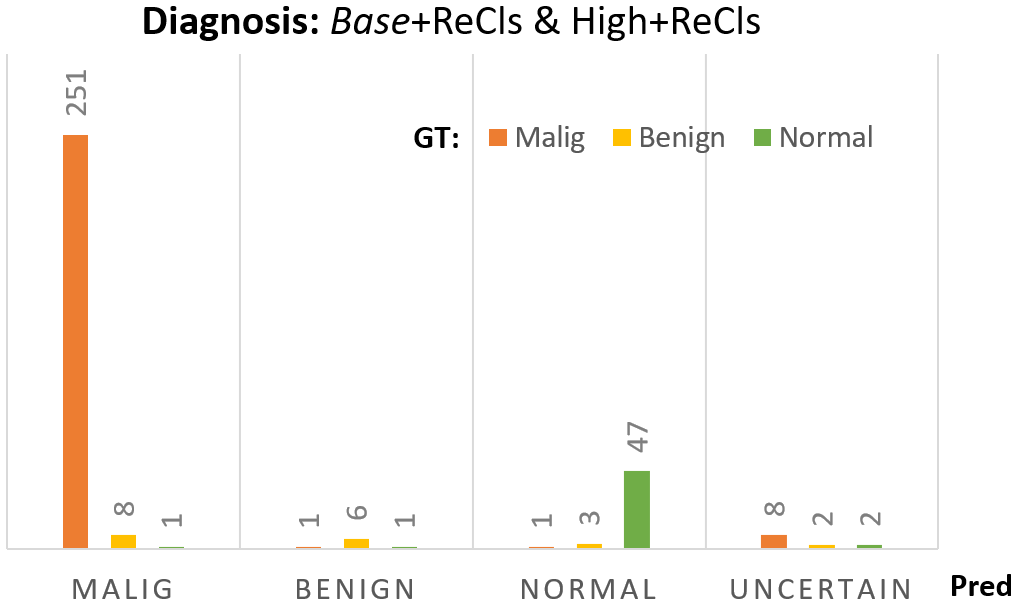}
	\hfill
	\caption{Patient-level malignancy classification of the Base+ReCls \& High+ReCls in the diagnosis scene. When the joint model classifies a patient as malignant, benign, or normal, most of the time, it is correct. When the joint predictions differ (\ie the UNCERTAIN set), the patient probably has some liver lesions.}
	\label{fig:Patient_level_chart}
\end{figure}

As shown in Table~\ref{tab:patient_level_malig}, the Base+ReCls \& High+ReCls has both high sensitivity and specificity in two different settings. A detailed prediction distribution for the diagnosis setting is in Fig.~\ref{fig:Patient_level_chart}. Only 3.6\% (12 cases) are labeled as uncertain by the joint model, and the rest predictions are mostly accurate. At the lesion level in Table~\ref{tab:DetSameLesionLevelDiagno}, the precision and recall improve slightly compared to High+ReCls in Table~\ref{tab:lesion_level_diagnosis}. It has a high ``rough-recall'' (ignoring the lesion type) for both the benign and malignant lesions (94.3\%).

\begin{table}[t]
	\caption{Lesion-level detection of High+ReCls model on Base+ReCls \& High+ReCls's consensus portion (\ie predict the same PatCls for each patient, 319 cases) in diagnosis setting. FN, FP, FL, and TP are the false negative, false positive, false label, and true positive lesion counts for each type using the ``match'' function in Equation~(\ref{eqn:matchGt}). Recall-rough measures the lesion recall rate ignoring the lesion type, \ie counting in both the TP and the FL. }
	\centering
	\hfill
	\scriptsize
\begin{tabular}{|c|c|c|c|c|c|c|c|}
\hline
\multirow{2}{*}{Type\textbackslash Metric} & \multicolumn{7}{c|}{High+ReCls} \\\cline{2-8}
                                          & FN   & FP   & FL   & TP    & Precision    & Recall     & Recall-rough      \\\hline
HCC                                       & 11   & 6    & 13   & 223   & 85.1\%   & 90.3\%   & 95.5\%    \\\hline
ICC                                     & 0    & 1    & 5    & 8     & 61.5\%   & 61.5\%   & 100\%   \\\hline
Meta                                      & 13   & 8    & 15   & 130   & 85.5\%   & 82.3\%   & 91.8\%    \\\hline
Hem                                     & 1    & 1    & 5    & 6     & 60.0\%   & 50.0\%   & 91.7\%    \\\hline
Other                                     & 8    & 6    & 24   & 25    & 65.8\%   & 43.9\%   & 86.0\%    \\\hline
Cyst                                      & 5    & 4    & 2    & 56    & 88.9\%   & 88.9\%   & 92.1\%    \\\hline
\textbf{All}                                       & 38   & 26   & 64   & 448   & 83.3\%   & 81.5\%   & 93.1\%    \\\hline
\textbf{Malig}                                     & 24   & 15   & 9    & 385   & 96.3\%   & 92.1\%   & 94.3\%  \\\hline

\end{tabular} 
	\hfill
	\label{tab:DetSameLesionLevelDiagno}
\end{table}

\subsection{Ablation}

\begin{figure}[t]
	\centering
    \hfill
	\includegraphics[width=0.98\linewidth]{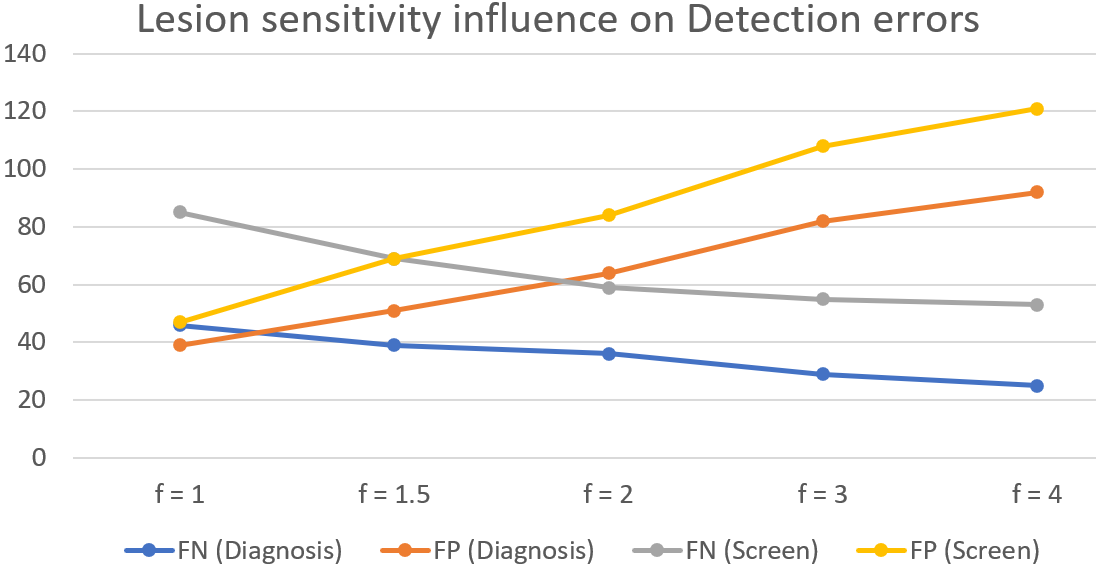}
	\hfill
	\caption{The influence of lesion sensitivity factor $f$ in Equation~(\ref{eqn:probscale}) on the detection results at the lesion level (without the ``reclassify'' module). A higher sensitivity leads to fewer miss detections (FN) at the cost of more false alarms (FP). This rule applies to both the diagnosis and the screening.}
	\label{fig:sensitivityScales}
\end{figure}
\subsubsection{Multi-sensitivity joint prediction: the advantage}
We conduct experiments with different sensitivities (i.e., change the scaling factor $f$) and show the impacts on the detection results in Fig~\ref{fig:sensitivityScales}. The lesion sensitivity has the opposite influence on FP and FN in both the screening and diagnosis settings. An optimal solution would maximize reliable information utilization while retaining essential entropy for the following human inspection. The complementary impacts of $f$ on Base+ReCls and High+ReCls demonstrate the necessity to harness both advantages and minimize disadvantages in the joint prediction for the final decision. 

\begin{figure}[t]
	\centering
    \hfill
	\includegraphics[width=0.98\linewidth]{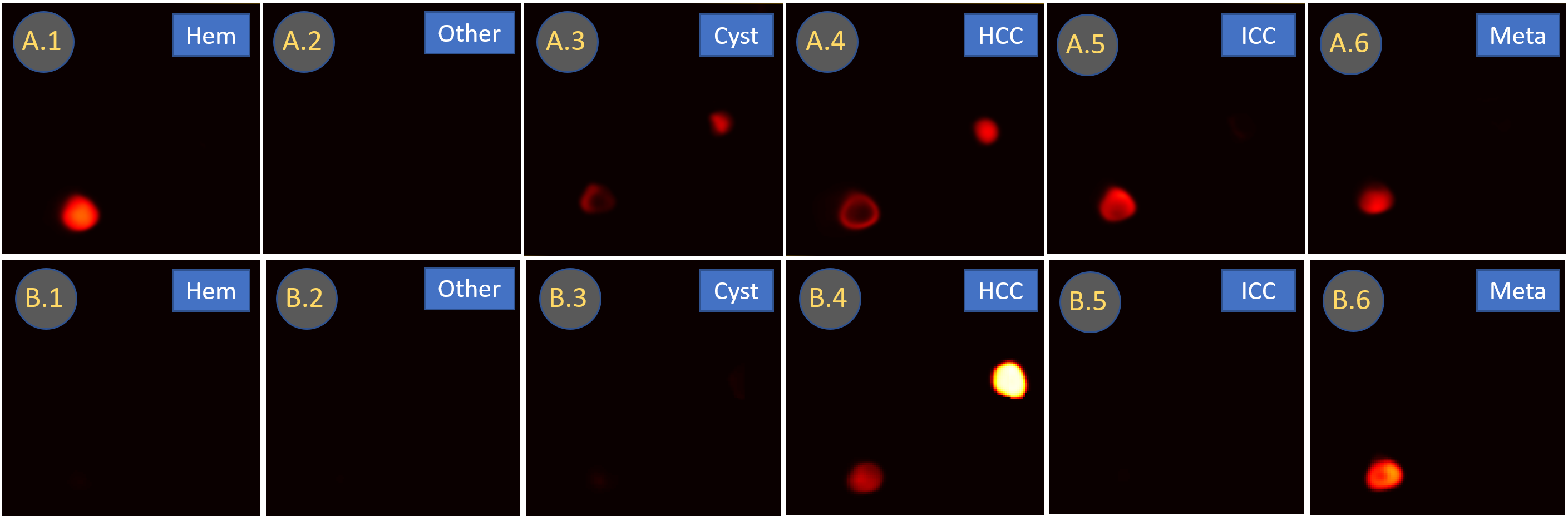}
	\hfill
	\caption{The individual probability maps from $UNet^{whole}$ (scaled $f=4$, A.1 -- A.6) and $UNet^{patch}$ (B.1 -- B.6) for the patch cropping in Fig.~\ref{fig:workflowPatch}. The dedicated $UNet^{patch}$ model generates lesion probability maps with better concentration and less ambiguity. }
	\label{fig:workflowProbMaps}
\end{figure}

\subsubsection{Analysis of $UNet$ probability maps}
The proposed workflow consists of two 3D \textit{UNet} models, $UNet^{whole}$ and $UNet^{patch}$. We study their generated probability maps for individual lesions in Fig.~\ref{fig:workflowProbMaps}, to compare the modeling characteristics. The input cropping in Fig.~\ref{fig:workflowPatch} contains two hard HCC tumors, and the multi-purpose $UNet^{whole}$ could not generate reliable results for either one. The (A.4) map has already been scaled by $f=4$, but it is still much weaker than lesions in the (B.4) map. The $UNet^{whole}$ generates signals in multiple probability maps (A.1, A.3, A.4, A.5, A.6) for the larger HCC lesion, showing poorer differentiating ability than the $UNet^{patch}$ model.

\begin{figure}[t]
	\centering
    \hfill
	\includegraphics[width=0.98\linewidth]{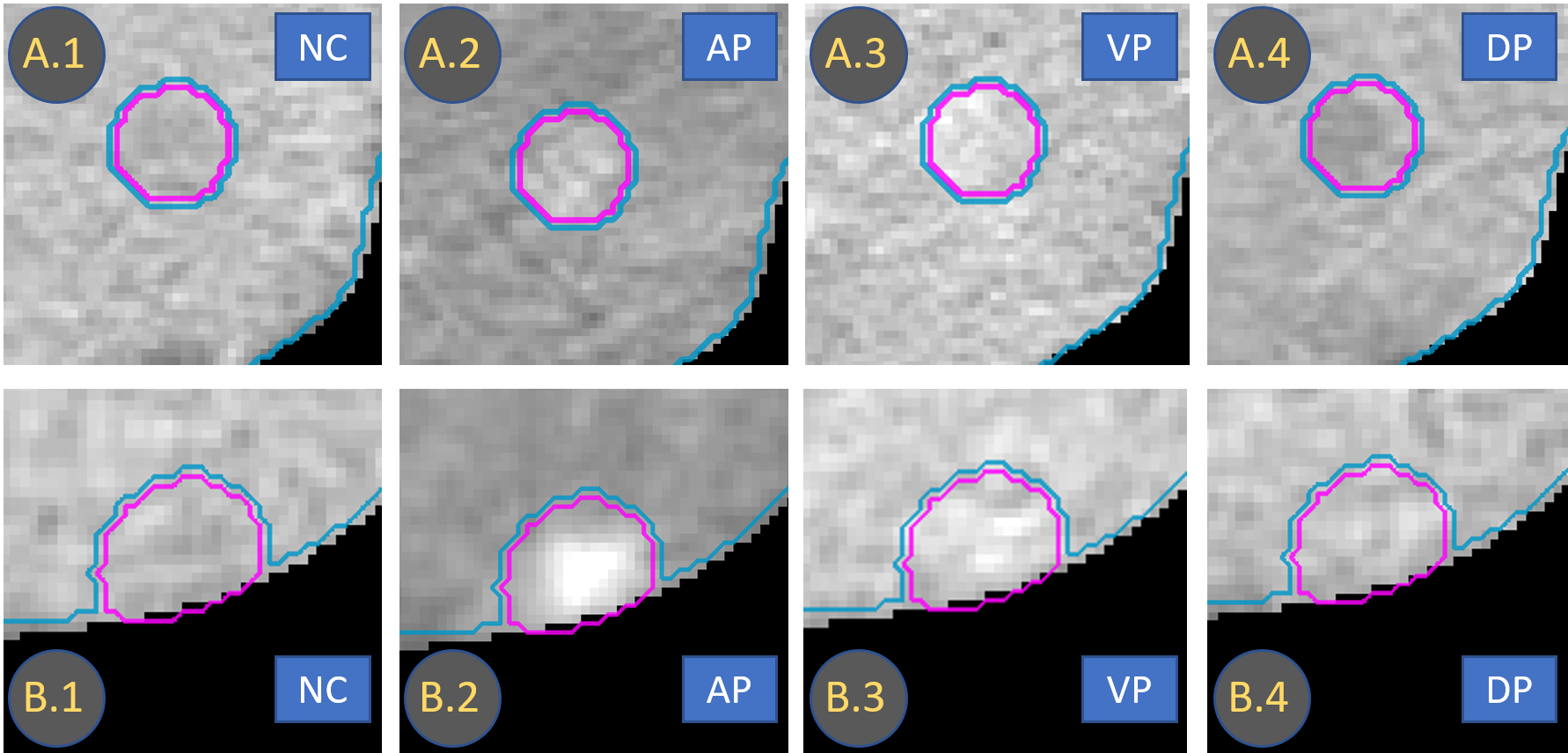}
	\hfill
	\caption{False positives in $UNet^{whole}$ results. The Case 02 cropping (A.1 -- A.4) and  Case 03 cropping (B.1 -- B.4) are both initially classified as HCC in $Det^{pred}_{f=4}$. They are removed in $Det^{ReCls}$ by $UNet^{patch}$ model through Equation~(\ref{eqn:reclassify}). }
	\label{fig:ablationF4L11FPsCorrected}
\end{figure}

\subsubsection{Lesion reclassification to reduce false positive detections}
The $Det^{pred}_{f=4}$ contains many FPs, illustrated in Fig.~\ref{fig:sensitivityScales} and Fig.~\ref{fig:ablationF4L11FPsCorrected}. Many FPs have suspicious appearances, such as venous phase hyper-enhancement similar to HCC, or vessel concentration similar to malignant lesions. But the liver is a large and complex organ, with many types of texture variations for various reasons. The $UNet^{whole}$ model not only segments the liver but also segments other related organs, which might limit its representation capability for liver lesions. The $UNet^{patch}$ in the ``reclassify'' module, on the other hand, only focuses on the liver region, with no interference from outside. With accurate organ masks from $UNet^{whole}$ to guide the cropping process, the training and testing patches (such as Fig.~\ref{fig:workflowPatch}) contain only the liver part.

\begin{figure}[t]
	\centering
    \hfill
	\includegraphics[width=0.98\linewidth]{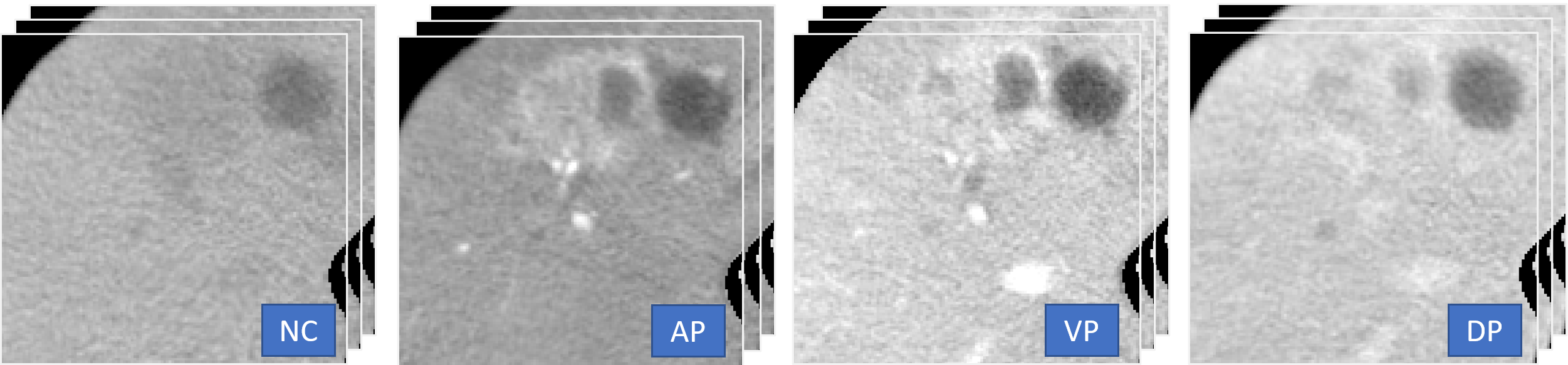}
	\hfill
	\caption{The Case 04 cropping is an ICC lesion, but it is initially mis-classified as HCC in $Det^{pred}_{f=4}$ in $UNet^{whole}$ results (\ie false label). }
	\label{fig:ablationF4L12FLsCorrected}
\end{figure}

\begin{figure}[t]
	\centering
    \hfill
	\includegraphics[width=0.98\linewidth]{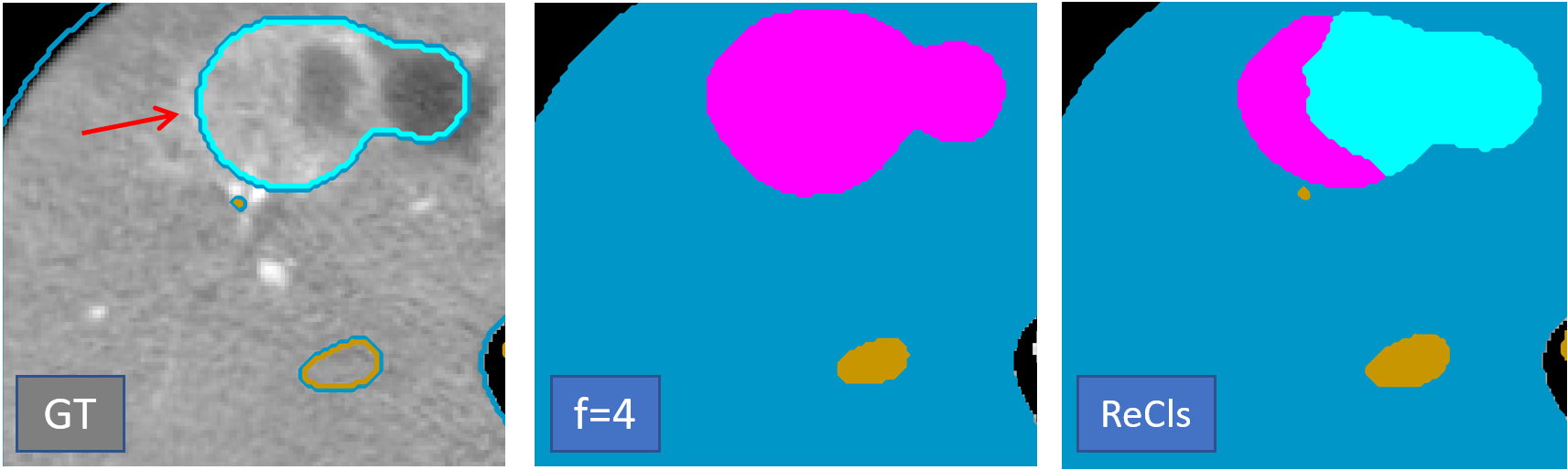}
	\hfill
	\caption{The ground truth mask, initial prediction mask (HCC), and the re-classified mask (ICC) for Case 04 cropping in Fig.~\ref{fig:ablationF4L12FLsCorrected}.}
	\label{fig:ablationF4L12FLsCorrected02}
\end{figure}

\subsubsection{Lesion reclassification to correct false labeling detections}
The $Det^{pred}$ contains some false label detections, where the model predicted a wrong label (lesion type) for a lesion. One example is in Fig.~\ref{fig:ablationF4L12FLsCorrected}, an ICC is predicted as HCC by the $UNet^{whole}$ model. With the help of the ``reclassify'' module, it is reclassified as ICC by $UNet^{patch}$ model through Equation~(\ref{eqn:reclassify}). In Fig.~\ref{fig:ablationF4L12FLsCorrected02} (ReCls), a smaller portion of the lesion is still classified as HCC due to the similarity to HCC outlooks~\cite{LIRADS}. In the end, the ``reclassify'' module will produce a single lesion label from volume-based voting. Similarly, many other FLs are corrected, leading to the precision increase in Base+ReCls compared to Base in Table~\ref{tab:lesion_level_diagnosis} and Fig.~\ref{tab:lesion_level_screen}.

\section{Discussion}
The workflow implementation is based on several foundations. Being familiar with UNet~\cite{Isensee2020nnUNetAS}~\cite{UNet} is necessary to understand the concepts of voxel-wise classification, individual probability maps, and segmentation mask. We develop the ``seg2det'' and ``match'' algorithms to generate precise lesion detections (with 3D masks) and pair-wise lesion matching, which enables subsequent lesion-shuffle training and reclassification. We customize the CT Labeler software for data curation, which improves labeling accuracy and efficiency.

Due to the complex conditions and various textures in the liver CT images, some lesions may not be able to be classified into a certain type by their image appearance. Therefore changing the lesion definition would influence the outcome. Moreover, due to the morphable nature of lesions as biological tissues, they may not have well-defined boundaries under certain circumstances. Therefore, the performance of deep learning models may rely on dataset characteristics, labeling protocols, and evaluation metrics. 

Our data is from a regional liver medical center whose population is limited to East Asian Han Chinese and the majority of records have liver-related diseases. In real-world screening, the majority of people have no liver tumors. Thus the performances for the diagnosis setting are more reliable than performances for the Screening setting in this paper. Nevertheless, the procedures of segmentation-for-detection formulation, coarse-to-fine classification, and context-aware augmentation would generalize to all lesion detection tasks.


\section{Conclusion}
In this paper, we propose a coarse-to-fine workflow for liver lesion detection in CT, which achieves high performance in both diagnosis and screening settings. By developing a customized multi-object labeling tool and adopting a semi-supervised learning approach, we curate a large-scale (1,631 patient cases) multi-phase CT image dataset, containing high-quality masks for seven organs\&vessels and six major types of liver lesions. Our workflow consists of a cascade of 3D segmentation models and procedures, and we describe them with detailed annotations and results. The proposed model could detect liver lesions with high precision and recall rates and classify the patient-level malignancy with high accuracy. We validate the proposed method in comprehensive experiments, with detailed lesion-level performances, practical patient-level analysis, and intuitive result visualization. Therefore the proposed model holds great clinical potential in both diagnosis and screening scenarios.

\ifCLASSOPTIONcaptionsoff
  \newpage
\fi





%



\bibliographystyle{IEEEtran}
\bibliography{mybibliography}

%











\end{document}